\documentclass{elsart} 
\usepackage{epsfig} 
\usepackage{psfig} 
\usepackage{amssymb} 
\usepackage{rotfloat} 
\def\cm{cm$^{-1}$} 
\begin{document}

\bibliographystyle{elsart-num} %
\title{Ab-initio density functional study of O on the Ag(001) surface} 
\author{ M.\ Gajdo\v s, A.\ Eichler and J.\ Hafner} 
\address{ Institut f\"ur Materialphysik and 
Center for Computational Materials Science\\ 
Universit\"at Wien, Sensengasse 8/12, A-1090 Wien, Austria 
\thanksref{email}} 
\thanks[email]{E-mail: marek.gajdos@univie.ac.at} 
\date{\today} 
 
\begin{frontmatter} 
\begin{abstract} 
The adsorption of oxygen on the Ag(100) is investigated by means 
of density functional techniques. Starting from a characterization 
of the clean silver surfaces oxygen adsorption in several 
modifications (molecularly, on-surface, sub-surface, Ag$_2$O) for 
varying coverage was studied. Besides structural parameters and 
adsorption energies also work-function changes, vibrational 
frequencies and core level energies were calculated for a better 
characterization of the adsorption structures and an easier 
comparison to the rich experimental data. 
\end{abstract} 
 
\end{frontmatter} 
 
\normalsize
\section{Introduction} 
\label{introduction} Despite the progress realized in surface 
science during the last decades, there are still unsolved 
questions for processes as elementary as atomic adsorption. One 
particular example is the adsorption of oxygen on the Ag(001) 
surface. 
 
Already an early work-function study \cite{ss:Engelhardt:57} 
indicates a non-trivial adsorption behavior: while the 
work-function increases monotonically with oxygen coverage for the 
(110) surface of silver, the work-function change for the (001) 
surface depends not only on coverage, but also on temperature. For 
surface temperatures higher than about 320 K oxygen exposure 
results always in an increased work-function, whereas at lower 
temperatures only small oxygen doses lead to a positive 
work-function change. For higher exposure the work-function 
decreases upon oxygen adsorption. Later this ``phase transition'' 
between a high- and a low-temperature structure could be 
reproduced and the different phases characterized by high 
resolution electron energy loss spectroscopy (HREELS) and 
low-energy electron diffraction (LEED) \cite{ssl:Fang:235}. 
Very recently these results were confirmed by 
a combined LEED, HREELS, x-ray photoemission spectroscopy (XPS) 
and x-ray photo-electron diffraction (XPD) study 
\cite{prb:Rocca:61}. Upon oxygen exposure several distinct species can 
be prepared (comp. table \ref{TabExp}): 
\begin{enumerate} 
\item Adsorption at low temperature ($T<150$K) leads to a 
 molecular species, which is 
stable up to 200~K and has been well characterized by 
XPD\cite{prb:Rocca:61}, HREELS\cite{bua97b} and thermal desorption 
spectroscopy (TDS)\cite{bua97a}. According to a scanning 
tunnelling microscopy (STM) study, the molecules adsorb in hollow 
sites, forming a c$(2\times4$) structure\cite{mes00}. \item After 
dissociation atomic oxygen adsorbs at the surface with an O 1s 
binding energy of 530.3~eV and a vibrational frequency of about 
30~meV. At around 190~K a sharp c$(2\times2)$ pattern appears and 
the frequency shifts up to 36~meV. An additional loss at 131~meV 
has been assigned recently on the basis of isotopically labelled 
oxygen to an electronic transition \cite{el:Benedek:53}. Based on 
XPD a missing-row p(2$\sqrt{2}\times\sqrt{2})$ reconstruction of 
the substrate was proposed (with the oxygen atoms forming a 
c($2\times2$) arrangement)\cite{prb:Rocca:61}. \item  Heating of 
this structure above $\sim320$~K lifts the reconstruction, the O 
1s binding energy changes from 530.3~eV to 528.3~eV and the 
vibrational frequency decreases again from 34-37~meV to 28-31~meV 
\cite{prb:Rocca:61}. \item Around the transition temperatures also 
an additional oxygen moiety appears with a binding energy of 
530.9~eV \cite{prb:Rocca:61}, which vanishes already at 350~K and 
could have been shadowed for lower temperatures by the dominant 
peak at 530.3~K. This species has been assigned to sub-surface 
oxygen. 
\end{enumerate} 
Interestingly only the higher temperature species characterized by 
an O 1s binding energy at 528.3~eV is active towards CO and 
C$_2$H$_4$ oxidation\cite{prb:Rocca:61}. 
Based on the structural models from XPD, a combined ab-initio and 
ultraviolet photoemission spectroscopy study was devoted to the 
investigation of electronic surface states \cite{ss:Savio:486}. In 
a recent STM experiment three different species of adsorbed oxygen 
($T_{\rm ads}$=~150~K) are reported for low coverage ($\Theta_{\rm 
O}\sim $0.1~ML) and assigned to hollow, bridge and on-top adsorbed 
atoms \cite{jcp:Schintke:114}. Finally, last year two density 
functional studies appeared investigating the oxygen adsorption on 
the Ag(001) \cite{ss:Cipriani:182,jcp:Wang:106}. Both studies are 
in a good agreement with our study. Also the adsorption on Ag(111) 
surface was investigated recently via DFT techniques 
\cite{prb:Li:65,prb:Li:67}.

\begin{table} 
\begin{tabular}{cccccc} 
\hline 
        & & T$~<~\sim200$~K &T$~<~\sim320$~K & \multicolumn{2}{c}{T$~>~\sim320$~K} \\ 
\hline 
$E_b$(O 1s)     & (eV)  & 532.2     &   530.3(530.9)  & 528.3         \\ 
$h\nu$          & (meV) & 30,84     &   34-37 (131)   & 28-31         \\ 
LEED/STM        &       & c(2$\times$4) &   c(2$\times$2) & p(1$\times$1) \\ 
$\Delta \Phi$   & (eV)  &   --  &   $\sim-0.2$    & $\sim0.3$   \\ 
$h_{\rm O}$     & (\AA) &   --  &   $-0.15$       & 0.6             \\ 
\hline 
\end{tabular} 
\vspace{0.4cm} 
 \caption{\label{TabExp}Experimental characterization of 
the different oxygen adsorption phases on Ag(001): O 1s binding 
energies $E_{b}$(O 1s), Ag-O stretching frequencies $h\nu$, LEED 
pattern, work-function change $\Delta \Phi$ and estimated height 
of O above the surface $h_{\rm O}$. The results for atomic oxygen 
are taken from Refs. 
\cite{ss:Engelhardt:57,ssl:Fang:235,prb:Rocca:61,chpl:Buatier:302,el:Benedek:53}, 
those for molecular oxygen (T$<~$200~K) from Refs. 
\cite{mes00,bua97b}.} 
\end{table} 
 
In the present study we use density functional theory (DFT) 
calculations to gain further insight into this complicated 
adsorption behavior. First we present our results for the clean 
silver surface. Then we discuss adsorption of oxygen in high 
symmetry (on-surface) sites of the (001) surface at different 
O-coverage. Subsequently results for molecular adsorption, the proposed missing-row 
reconstruction and the silver oxide surface are presented. 
The paper closes with a discussion of our results in 
the light of the experimental data. 
 
\section{Methodology} 
\label{Methodology} The calculations are performed using the 
Vienna ab-initio simulation package VASP \cite{vasp,prb:Kresse:54} 
which is a DFT code, operating in a plane-wave basis set. The 
electron-ion interaction is described using the 
projector-augmented-wave (PAW) method 
\cite{prb:Blochl:50,prb:Kresse:59} with plane waves up to an 
energy of $E_{\rm cut}=~250$~eV. For exchange and correlation the 
functional proposed by Perdew and Zunger \cite{prb:Perdew:23} is 
used, adding (non-local) generalized gradient corrections (GGA) 
following Perdew et al. \cite{prb:Perdew:46}. If not mentioned 
differently we have used a ($12\times12\times1$) k-points mesh for 
the integration over the Brillouin zone for the p(1$\times$1) 
cell, (6$\times$6$\times$1) for 
c(2$\times$2),(4$\times$4$\times$1) for p(2$\times$2) and a 
(3$\times$3$\times$1) mesh for a p(3$\times$3) cell. 
 
The system itself is modelled by six layers of Ag separated by 14 
\AA~ of vacuum with oxygen adsorbed on one side of the slab. The 
upper two layers of the Ag surface are allowed to relax, the 
remaining atoms are fixed at their ideal bulk positions ($a_{\rm 
GGA}$=4.16~\AA, $a_{\rm exp}$=4.08~\AA). 
 
{ The free molecule  is characterized by a calculated stretch 
frequency of 1561 \cm at an equilibrium bond length of 1.23 \AA. 
The corresponding experimental values are 1580 \cm and 1.21 \AA 
\cite{hub79}. The problematic description of the O$_2$ binding 
energy (E$_{O_2}^{GGA}$=6.25 eV, E$_{O_2}^{exp}$=5.23 eV) stems 
mainly from the error in the energy of the free atoms, where high 
density gradients make an accurate description more difficult. So 
the reported energies (with respect to the calculated binding 
energy of molecular oxygen) should be of much higher accuracy. 
This is for example reflected by the calculated formation energy 
for Ag$_2$O of 0.39 eV compared to the literature value of 0.32 
eV\cite{weast}.} 
 
Adsorbate-substrate stretching frequencies have been obtained from 
finite displacements of the oxygen atoms along all three Cartesian 
directions ($\Delta=\pm 0.04$~\AA). In order to increase the 
accuracy we performed the calculations with a harder 
pseudo-potential requiring an energy cut-off of $E_{\rm 
cut}$=~400~eV. Diffusion barriers were determined with the nudged 
elastic band method \cite{ss:Mills:324,jcp:Ulilitsky:92}. 
 
Core-level shifts were calculated 
in a ($2\sqrt{2}\times\sqrt{2}$) cell sampled by a  ($3\times6\times1$) k-points 
mesh. Further details concerning  the calculation of the core levels (final state effects) 
are given in Ref. \cite{koe02}.

\section{Results} 
\subsection{Clean Ag(001)} 
\label{Clean Ag(001)} The clean Ag(001) surface is well 
characterized, experimentally as well as theoretically 
 \cite{ss:Cipriani:182,jcp:Wang:106,prb:Smith:22,ssr:Bohnen:19,prb:Methfessel:46,prb:Quinn:43,ss:Erschbaumer:243,prb:Giesen:35}. 
Calculations have been done for slabs with 4 to 8 layers, to check 
for errors due to the finite thickness of the Ag film. Both 
surfaces of the slab were relaxed. Table \ref{clean} compiles our 
results for inter-layer relaxation, surface energy and 
work-function together with values from the literature. We find 
that structural, as well as electronic parameters agree well with 
earlier  results.  A slight inward relaxation of the surface plane 
is compensated by outward relaxations of the lower layers of 
comparable size. The surface energy is underestimated by about 30 
\% compared to earlier LDA calculations and experiments, a 
property of the GGA that has already been discussed earlier 
\cite{ss:Vitos:411}. However, since it is well established that 
the GGA leads to a more reliable description of the adsorption 
process we did our calculations with the inclusion of gradient 
corrections. 
 
\begin{table} 
\begin{center} 
\begin{tabular}{ccccccccc} 
\hline & & \multicolumn{3}{c}{Present work(GGA)} & Theory 
(LDA) & (GGA)$^c$ & (GGA)$^d$ & Exp. \\ 
& & \multicolumn{3}{c}{ Number of layers} & & & & \\ 
& & 4 & 6 & 8 & & &  & \\ 
$\Delta d_{12}$&[\%] & -1.9 & -1.6 & -1.5 & -1.3$^{a}$ & -2.0 
& -1.7 & 0$\pm$1.5$^{e}$\\ 
$\Delta d_{23}$&[\%] & 0.3 & 0.7 & 0.8 & 1.0$^{a}$ & 0.8 
& 0.7 & 0$\pm$1.5$^{e}$\\ 
$\Delta d_{34}$& [\%] & - & 0.8 & 0.9 & 0.8$^{a}$ & - 
& 0.2 & - \\ 
$\sigma$ &[J/$m^{2}$] & 0.82 & 0.80 & 0.78 & 1.21$^{a}$& 0.8 
& - & 1.27$^{f}$ \\ 
$\Phi$ &[eV] & 4.22 & 4.38 & 4.44 & 4.43$^{b}$ & 4.62 
& 4.33 & 4.42$^{g}$ \\ 
\hline 
\end{tabular} 
\vspace{0.4cm} 
\end{center} 
\caption{Inter-layer relaxation ($\Delta d_{12}$, $\Delta d_{23}$, 
$\Delta d_{34}$) in \%, surface energy ($\sigma$) in J/m$^2$ and 
work-function ($\Phi$) in eV for a clean Ag(001) surface as a 
function of slab thickness. Earlier theoretical and experimental 
results are taken from Refs. \cite{ssr:Bohnen:19} 
(a),\cite{prb:Methfessel:46} (b), \cite{ss:Cipriani:182} (c), 
\cite{jcp:Wang:106} (d) and \cite{prb:Quinn:43} (e), 
\cite{ss:Erschbaumer:243} (f), \cite{prb:Giesen:35} 
(g).}\label{clean} 
\end{table} 
 
Concerning the slab thickness dependence, results for the six 
layer slab, which were used for the adsorption studies are 
converged within 0.1\% (0.02 \AA) for the inter-layer relaxation, 
within 30 mJ/m$^2$ for surface energies and the work-function 
varies only by 60~meV when going from six to eight layers. 
 
\subsection{Adsorption in high symmetry on-surface sites} 
\label{Adsorption on high symmetry sites} As a first step towards 
a deeper understanding of the adsorption process, we put oxygen at 
varying coverage ($\Theta=\frac19, \frac14, \frac12, \frac34, 1$ 
~ML) in the high symmetry positions (top, bridge, hollow) of the slab, 
optimized the geometry and calculated frequencies and 
work-function changes. For the characterization of the energy 
balance during adsorption we chose two slightly distinct 
quantities. First, the adsorption energy per adsorbed atom 
($E_{\rm ads}$) with respect to half an oxygen molecule defined as 
\begin{equation} 
E_{\rm ads}=(E_{\rm tot}-E_{\rm clean}-\frac{N_{\rm O}}{2}\cdot 
E_{\rm O_2})/ N_{\rm O}, 
\end{equation} 
\label{adsorbtion energy} with the total energy of the adsorption 
system (containing $N_{\rm O}$ oxygen atoms) $E_{\rm tot}$, the 
energy of the clean surface ($E_{\rm clean}$) and the energy of a 
free O$_2$ molecule $E_{\rm O_2}$. Following this definition, 
$E_{\rm ads}$ decreases (i.e. becomes more exothermic) with 
increasing Ag-O bond strength. Secondly, we use a (generalized) 
surface energy per area $A$ ( especially for the comparison with the 
reconstructed missing-row structure, discussed below) 
\begin{equation} 
\label{surface energy} \sigma = (E_{\rm tot}-N_{\rm Ag}\cdot 
E_{\rm Ag}-\frac{N_{\rm O}}{2}\cdot E_{\rm O_2})/A-\sigma_{\rm 
clean}, 
\end{equation} 
 where $N_{\rm Ag}$ and $E_{\rm Ag}$ are the 
number of silver atoms per cell and the total energy of a bulk 
silver atom, respectively. The subtraction of the surface energy of a 
clean silver surface $\sigma_{\rm clean}$ accounts for the second 
(uncovered) surface of the slab. This surface energy describes the 
energy cost for creating a surface minus the energy gain due to 
oxygen adsorption. 
 
\begin{figure}[htb] 
\centerline{\psfig{figure=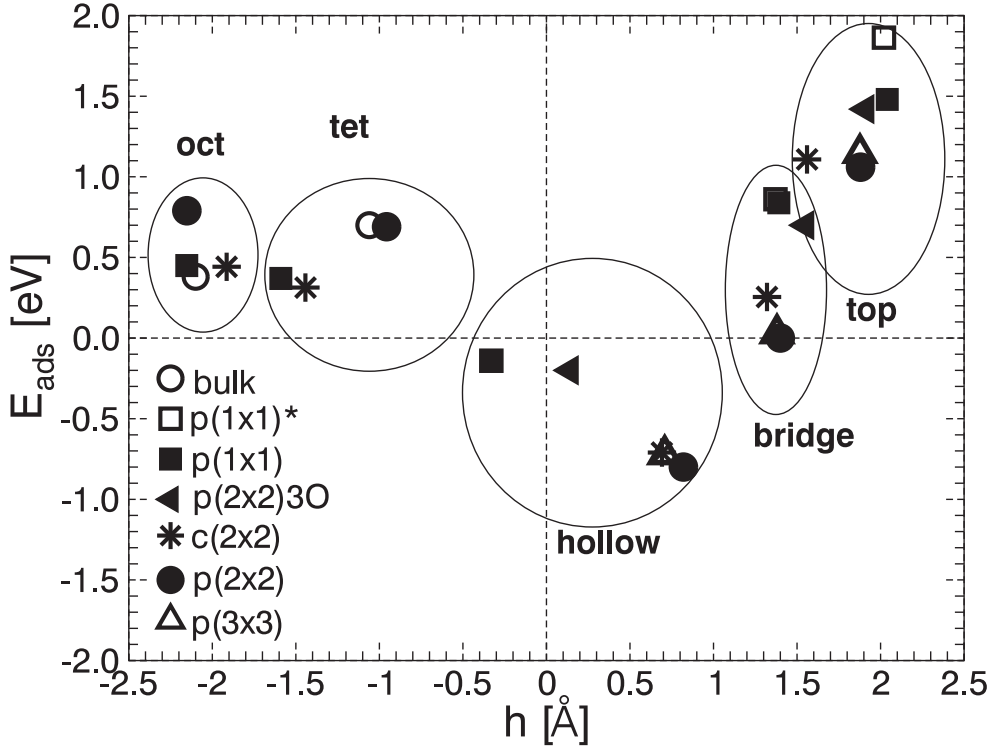,width=13cm,clip=true}} 
\nopagebreak \caption{\label{FigSites} Adsorption energy 
($E_{ads}$) versus height of the adsorbed O atom 
 for various adsorption sites and coverage. The 
second data set for full coverage (denoted by an asterisk) has 
been obtained from a spin polarized calculation. For subsurface 
adsorption also energies for incorporation in bulk Ag are 
included (empty circles).} 
\end{figure}

The results for the various adsorption sites at different coverage 
are compiled in table \ref{surface}; fig.\ref{FigSites} gives a 
graphical summary of the adsorption energy versus adsorption 
height. At every coverage, adsorption in hollow sites is favored 
while on-top and bridge adsorption is nearly always endothermic. 
Only at a coverage of $\Theta=0.25$~ML adsorption in bridge sites 
becomes thermoneutral, but remains still $\approx 0.8$~eV less 
favorable compared to adsorption in the hollow sites. { All these 
findings for the adsorption energetics are in nice agreement with 
two recent DFT studies \cite{ss:Cipriani:182,jcp:Wang:106}. This 
strong preference of hollow adsorption} makes the interpretation 
of the STM images in Ref. \cite{jcp:Schintke:114} in terms of 
bridge- and top-adsorbed meta-stable oxygen species very 
improbable, unless there were huge barriers between the adsorption 
sites. This point will be addressed in the following section. 
 
At full coverage we performed additionally spin-polarized 
calculations. For atop adsorption a substantial magnetic moment of 
the O atom of 1.4~$\mu_{\rm B}$ persists, leading to a decrease in 
the adsorption energy by -0.4~eV at an unchanged geometry. In the 
higher coordinated sites the effect of spin polarization is 
negligible, due to the stronger interaction with the substrate. At 
lower coverage this effect is expected therefore to be even 
smaller due to stronger Ag-O interaction, justifying the neglect 
of spin polarization in further calculations. 
 
The inward relaxation of the first substrate layer, which was 
observed for the clean surface, persists up to a coverage of 
$\Theta=\frac14$~ML; for higher coverage it is gradually turned 
into an expansion. For the second inter-layer spacing the sign of 
the relaxation changes already at $\Theta=\frac14$~ML. The 
adsorption heights for the three high symmetry adsorption sites 
vary such, that the Ag-O bond length remains always between 1.9 
and 2.3~\AA. This results in adsorption heights of around 2.0~\AA, 
1.4~\AA~and 0.7~\AA~for top, bridge and hollow adsorption, 
respectively. The correlation between adsorption energy and 
adsorption height reveals a very peculiar character of 
O-adsorption in the hollow position (see fig. \ref{FigSites}). 
Whereas for bridge- and top-adsorbed O the adsorption-height is 
almost independent of the coverage and the adsorption energy 
becomes more endothermic as the coverage increases, 
hollow-adsorbed O sinks deeper into the surface with increasing 
coverage - at full coverage the stable adsorption sites are 
located even below the surface-plane ($h_{\Theta=1}=-0.33$~\AA~). 
This reduction of the adsorption-height is accompanied by a strong 
outward relaxation of the top Ag-layer ($\Delta d_{12}$=31\%). 
This expanded adsorption structure may be interpreted as the 
formation of an silveroxide layer which is weakly bonded to the 
Ag-substrate. 
A similar behavior is found already for $\Theta=\frac34$~ML 
(p(2$\times$2)3O). In this structure two inequivalent adsorption 
sites exist for oxygen. Two out of three oxygen atoms are 
equivalent due to symmetry and stay above the surface at a height 
of $h=0.26$~\AA, while  the third oxygen atom sinks below the 
first layer ($h=-0.09$~\AA~). 
 
The work-function change $\Delta\Phi$ for top and bridge adsorbed 
oxygen increases monotonically with coverage. For the hollow site 
this trend holds only up to $\frac{1}{2}$~ML, from where on 
$\Delta\Phi$ decreases again, due to the aforementioned negative 
adsorption heights at 0.75~ML and 1~ML. 
 
The vibrational frequency perpendicular to the surface exhibits 
for all three adsorption sites an increasing trend with coverage up to 
around 0.5~ML. From there on the vibrational modes become softer again due 
to the pronounced weakening of the Ag-O bond at high coverage.

\begin{sidewaystable} 
\begin{center} 
\begin{tabular}{r|rrr|rrr|rrr|rrr|rrr} 
 \hline 
 & \multicolumn{3}{c}{p(1$\times$1)O}& 
 \multicolumn{3}{c}{p(2$\times$2)3O} & \multicolumn{3}{c}{c(2$\times$2)O} & 
 \multicolumn{3}{c}{p(2$\times$2)O} & \multicolumn{3}{c}{ p(3$\times$3)O}\\ 
 & \multicolumn{3}{c}{$\Theta=1$~ML}& 
 \multicolumn{3}{c}{$\Theta=0.75$~ML} & \multicolumn{3}{c}{$\Theta=0.5$~ML} & 
 \multicolumn{3}{c}{$\Theta=0.25$~ML} & \multicolumn{3}{c}{$\Theta=0.11$~ML}\\ 
    & t & b & h & t & b & h & t & b & h & t & b & h & t & b & h \\ 
\hline $\sigma$ [J/$m^{2}$] & 3.54 & 2.33 & 0.52 & 2.72 & 
 1.73 & 0.48 & 1.83 & 1.04 & 0.15 & 1.24 & 0.75 & 0.38 & 1.00 & 0.77 & 0.62 \\ 
$E_{ads}$ [eV] & 1.48 & 0.83 &--0.14 & 1.42 & 0.70 &--0.20 & 1.11 
& 0.26 &--0.71 & 1.06 & -0.00 & --0.80& 1.14 & 0.03 & --0.72 \\ 
$h$ [\AA] & 2.04 & 1.39 &--0.33 & 1.91 & 1.55 & 0.14 & 1.56 & 1.32 
& 0.69 & 1.88 & 1.40 & 0.81 & 1.88 & 1.38 & 0.71 \\ 
$\Delta d_{12}$ [\%] &--1.1 &--0.4 & 31.1 & --0.8 &--1.5 & 14.5 & 
6.4 & --1.9 & 4.8 & --1.4 & --1.2 & --0.8 & --1.6 & --1.7 & --1.6 \\ 
$b_{1}$ [\AA] & 0 & 0 & 0 & 0.22 & 0.08 & 0 & 0.77 & 0 & 0 & 0.13 
& 0.07 & 0 & 0.04 & 0.06 & 0.03 \\ 
$\Delta d_{23}$ [\%] & 0.2 &--0.3 &--0.4 & -0.4 & 2.8 &--1.6 & 
--1.9 & --1.3 & --0.8 &--0.1 & 0.5 & 
--1.0 & 0.6 & 0.1 & 1.4 \\ 
$b_{2}$ [\AA] & 0 & 0 & 0 & 0 & 0.08 & 0.1 
 & 0 & 0 & 0.03 & 0 & 0.01 & 0.14 
 & 0.02 & 0.03 & 0.11 \\ 
$d_{Ag-O}$ [\AA] & 2.04 & 2.02 & 2.11 & 2.00 & 2.09 & 2.10 & 1.95 
& 2.05 & 2.19 & 1.96 & 2.07 & 2.26 & 
1.97 & 2.07 & 2.26 \\ 
$\Delta\Phi$ [eV] & 4.44 & 4.04 &--0.14 & 3.91 & 3.29 & 0.47 & 
3.30 & 3.22 & 0.94 & 1.96 & 1.69 & 
0.68 & 1.18 & 0.87 & 0.34 \\ 
$\nu$ [meV] & 53.5 & 52.8 & 23.3 & 51.5 & 45.2 & 26.1 & 67.6 & 51.3 & 29.4 & 64.6 & 50.0 & 30.8 & 63.0 & 48.0 & 24.2 \\ 
 \hline 
\end{tabular} 
 \caption{Results for oxygen adsorbed in high symmetry sites (t 
- top, b - bridge, h - hollow) on Ag(001) for coverages varying 
between $\Theta=0.11$~ML and $\Theta=1$~ML. The reported 
quantities are: $\sigma$ - surface energy, $E_{ads}$ - adsorption 
energy, $h$ - height of O with respect to the average surface 
plane atoms, $\Delta d_{12}$, $\Delta d_{23}$ - change of the 
average inter-layer spacing, $b_{1}$,$b_{2}$ - buckling of 
$1^{st}$ and $2^{nd}$ layer, $d_{Ag-O}$ - minimal Ag-O distance, 
$\Delta\Phi$ - oxygen induced work-function change, $\nu$ - oxygen 
frequency perpendicular to the surface.} \label{surface} 
\end{center} 
\end{sidewaystable} 
 
\subsection{Transition between high-symmetry sites - surface diffusion} 
\begin{figure}[htb] 
\centerline{\psfig{figure=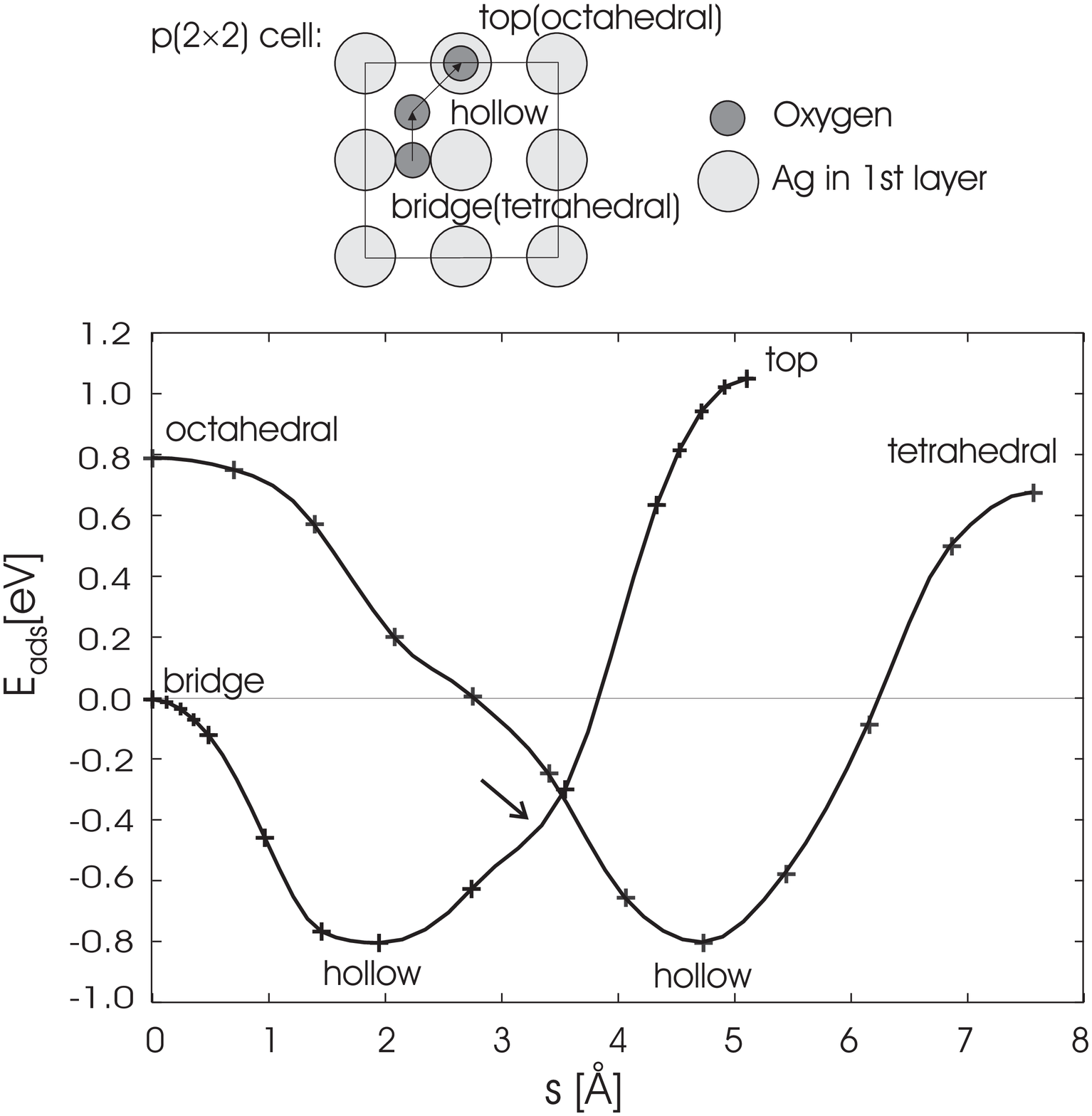,width=12cm,clip=true}} 
\nopagebreak \caption{\label{FigPath} Adsorption energy $E_{ads}$ 
along a pathway connecting the adsorption sites bridge - hollow - 
top calculated in a $2\times2$ cell (octahedral - hollow - 
tetrahedral sites for the pathway leading into subsurface 
positions; compare sketch). Calculated points are indicated by 
crosses, the connecting lines are spline fits, based on forces.} 
\end{figure}

In a recent low temperature STM study \cite{jcp:Schintke:114} 
three different oxygen species have been imaged for low oxygen 
coverage. These have been assigned to  hollow adsorbed 
oxygen atoms and two meta-stable species residing in bridge 
and on-top sites, respectively.  Considering the large differences in 
the adsorption energies for these three sites, this could only be 
possible if there were significant barriers separating the 
meta-stable sites from the hollow position. In order to clarify 
this point we performed nudged-elastic-band (NEB) 
\cite{ss:Mills:324,jcp:Ulilitsky:92} calculations to determine the 
diffusion barriers between neighboring high symmetry sites. Fig. 
\ref{FigPath} shows the variation of the potential energy for the 
movement of an oxygen atom from a bridge-site via hollow towards 
the on-top position, following the pathway shown in the inset. For 
this calculation (performed in a p(2$\times$2) cell) again the upper 
two silver layers were allowed to relax. 
 
The results show that bridge and on-top sites are only 
saddle-points of the potential energy surface for this coverage. 
Hence on the basis of these findings the interpretation given in 
the STM work becomes very improbable. However, the calculation 
reveals another possible (3-fold)  adsorption site between 
hollow and on-top, reflected by a dip 
along the pathway (arrow in the graph on fig. \ref{FigPath}). 
For a further investigation we put 
the oxygen atom artificially at that position and performed again 
a structural optimization. Our calculations did not lead to any 
local minimum. 
 
\subsection{Sub-surface adsorption} 
\label{Sub-surface adsorption} We have also considered the 
sub-surface adsorption of oxygen, since such species are believed 
to be important for the selectivity of silver in catalyzing 
ethylene epoxidation. \cite{jpch:Hoek:93} We investigated the 
sub-surface adsorption for coverages between a quarter and full 
monolayer (p(2$\times$2), c(2$\times$2) and p(1$\times$1) 
structures) in the octahedral and tetrahedral sites, the results 
of which are compiled in table \ref{sub-surface} and 
fig.\ref{FigSites}. { For comparison, we included in fig. 
\ref{FigSites} also the energies for oxygen incorporated in the 
tetrahedral and octahedral sites of bulk Ag (in a 
$3\times3\times3$ cell). Sub-surface adsorption is an endothermic 
process compared to molecular O$_2$ at all investigated coverages 
(see fig. \ref{FigSites}). However, there are interesting trends 
of the adsorption energy in coverage and in comparison to the 
dilute solution of oxygen in the bulk. In contrast to the 
on-surface adsorption the variation of the adsorption energy with 
coverage is non-monotonous, the smallest (least endothermic) 
energy is calculated for $\Theta=0.5$ ML, i.e. at a composition of 
the surface layer corresponding to stoichiometric Ag$_2$O. And the 
dilute solution of oxygen in bulk silver is an extremum for both 
adsorption sites: for the octahedral site it is more stable than 
the high coverage subsurface case, whereas the tetrahedral 
interstitial is even less stable than the quarter coverage 
subsurface position. 
 
This peculiar behavior can be understood when 
considering the geometries of the two interstitial sites. Placing 
an oxygen atom into an ideal (unrelaxed) Ag bulk leads to Ag-O 
bond lengths of 1.80 \AA~ for the tetrahedral and 2.08 \AA~ for 
the octahedral site. So compared to the typical Ag-O bond lengths 
of 2.0-2.2~\AA~ (compare table \ref{surface}) the octahedral site 
can easily accommodate an oxygen atom, whereas the tetrahedral 
site is too small. This explains already the strong preference for 
the octahedral site in the dilute solution of oxygen in the bulk. 
At the surface the incorporation of oxygen induces pronounced 
geometrical changes, partially just for sterical reasons, so that 
the site becomes large enough to accommodate the additional atom; 
partially because the formation of strong Ag--O bonds weakens the 
Ag-Ag bonds and leads hence to an outwards relaxation of the first 
layer. 
 
For the larger octahedral site, the second contribution dominates: 
already at the lowest coverage (p(2$\times2$)), the outward 
relaxation together with the up-shift of the immediate silver 
neighbor in the first layer (reflected by a large buckling of 
0.42~\AA), expands the octahedral site such that two of the six 
Ag--O bonds increase beyond 2.3~\AA~ and are significantly 
weakened. This results in the weakest subsurface adsorption energy 
of this study. With increasing coverage, the outward relaxation 
increases dramatically, so that a slightly different adsorption 
geometry between the first and second layer becomes available. In 
this position the Ag--bond to the third layer is now completely 
absent ($d_{Ag-O}^{l}>3$~\AA), but for that the remaining 5 bonds 
are stronger and make this position lower in energy. 
 
The smaller tetrahedral site on the other hand has to rely on the 
(local) expansion of the substrate. At low coverage the two silver 
neighbors in the first layer have to shift up significantly 
(leading to a buckling of 0.64~\AA), so that the interstitial 
becomes large enough. Such a large distortion costs energy and 
makes the tetrahedral site unfavorable at low coverage. With 
increasing coverage, the weakening of the Ag--Ag bonds opens up 
the first interlayer spacing, so that adsorption in the 
tetrahedral site becomes not only possible, but even more 
favorable than adsorption in the octahedral site.} 
 
\begin{table} 
\begin{center} 
\begin{tabular}{r|rr|rr|rr} 
 \hline 
& \multicolumn{2}{c}{p(1$\times$1)} & \multicolumn{2}{c}{c(2$\times2$)} & \multicolumn{2}{c}{p(2$\times$2)}\\ 
& \multicolumn{2}{c}{$\Theta=1$~ML} & \multicolumn{2}{c}{$\Theta=0.5$~ML} & \multicolumn{2}{c}{$\Theta=0.25$~ML}\\ 
                      & oct & tet & oct & tet & oct & tet \\ 
 \hline 
$\sigma$ [J/$m^{2}$]    & 1.65  & 1.49  & 1.24  & 1.12  & 1.13  & 1.08\\ 
$E_{ads}$ [eV]          & 0.46  & 0.37  & 0.44  & 0.31  & 0.79  & 0.69\\ 
$\Delta d_{12}$ [\%]    & 50.4  & 54.2  & 35.6  & 44.9  & 8.7   & 16.0\\ 
$b_{1}$ [\AA]           & 0     & 0     & 0.58  & 0     & 0.42  & 0.64\\ 
$\Delta d_{23}$ [\%]    & 0.9   & --0.9 & 1.4   & 0.6   & 3.0   & --0.5\\ 
$b_{2}$ [\AA]           & 0     & 0     & 0     & 0     & 0     & 0.12\\ 
$d_{Ag-O}^{u}$ [\AA]    & 2.15  & 2.17  & 2.21  & 2.15  & 2.38  & 2.10\\ 
$d_{Ag-O}^{m}$ [\AA]    & 2.30  & -     & 2.27  & -     & 2.23  & -\\ 
$d_{Ag-O}^{l}$ [\AA]    & 3.07  & 2.19  & 3.02  & 2.20  & 2.32  & 2.18\\ 
$\Delta\Phi$ [eV]       & 0.15  & 0.37  & 0.15  & 0.37  & 0.27  & 0.15 \\ 
$\nu$ [meV]             & 34.2  & 37.4  & 36.0  & 48.8  & 29.5  & 49.6 \\ 
 \hline 
\end{tabular} 
\end{center} 
\vspace{0.4cm} \caption{Results for oxygen adsorbed in the 
octahedral (oct) and tetrahedral (tet) sub-surface sites at 
coverages of 1~ML (p(1$\times$1)), 0.5~ML (c(2$\times$2)) and 
0.25~ML (p(2$\times$2)). The reported quantities are: $\sigma$ - 
surface energy,$E_{ads}$ - adsorption energy, 
 $\Delta d_{12}$, $\Delta d_{23}$,- change of the average 
inter-layer spacing, $b_{1}$,$b_{2}$ - buckling of $1^{st}$ and 
$2^{nd}$ layer, $d_{Ag-O}^{u}$, $d_{Ag-O}^{m}$, $d_{Ag-O}^{l}$ - 
shortest Ag-O distance to the upper, middle and lower Ag layer, 
$\Delta\Phi$ - the work function change compared to the clean 
Ag(001) surface and the oxygen frequency perpendicular to the 
surface $\nu$.}\label{sub-surface} 
\end{table} 
 
 The vibrational frequencies perpendicular to the surface for the 
 tetrahedral and octahedral site are remarkably similar to those for the 
corresponding on-surface sites, bridge and hollow, respectively. 
 
Similar to the on-surface case we calculated the potential energy 
along a pathway connecting the subsurface sites with the 
on-surface hollow position in a p($2\times2$) cell. The monotonic 
decrease in energy towards the hollow position demonstrates that 
isolated sub-surface sites are highly unstable at this coverage 
and can only be stabilized by co-adsorbed on-surface oxygen, 
preventing the migration back to the surface.

\subsection{Missing-row reconstruction} 
\label{Missing-row reconstruction} 
 
For the low-temperature phase (below 320 K, compare table 
\ref{TabExp}), a missing-row reconstructed phase with a 
p($2\sqrt2\times\sqrt2$) periodicity has been proposed by Rocca 
{\em et al.} on the basis of XPD experiments \cite{prb:Rocca:61}. 
The still rather regular position of the oxygen atoms should give 
rise to the observed c(2$\times$2) LEED pattern 
\cite{ssl:Fang:235}.

\begin{figure}[htb] 
 \centerline{ \psfig{figure=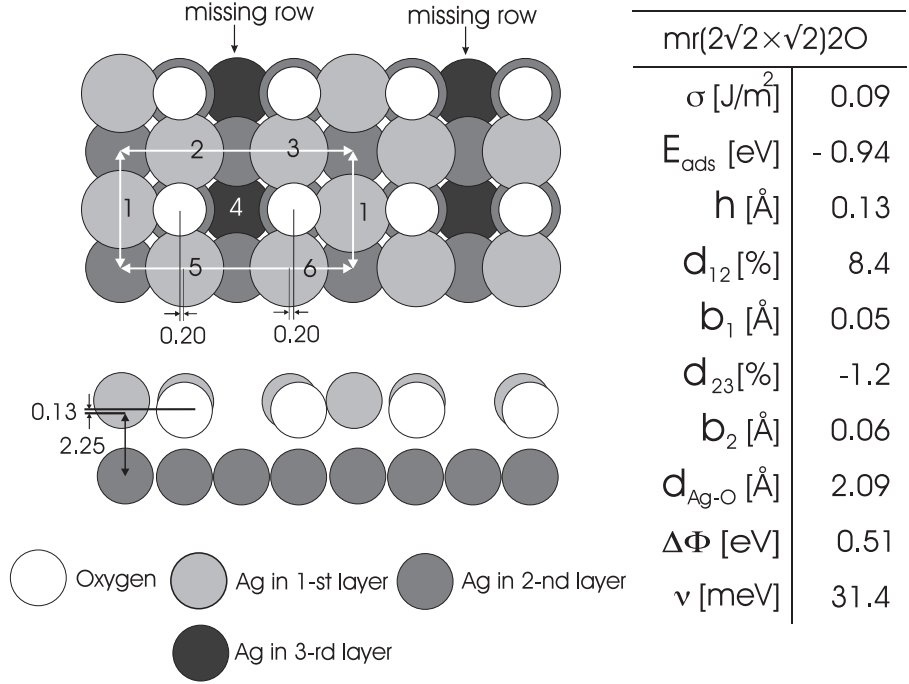,width=12cm,clip=true} } 
\nopagebreak \caption{\label{FigSketch} Results for the 
($2\sqrt{2}\times\sqrt{2}$) missing-row reconstruction with two 
oxygen atoms in the pseudothreefold hollows. The sketch shows a 
top and side view of the optimized structure (all distances 
in~\AA). The reported quantities in the table are: $\sigma$ - 
surface energy, $E_{ads}$ -  adsorption energy with respect to a 
clean reconstructed surface, $h$ - height of O with respect to the 
average surface plane, $\Delta d_{12}$, $\Delta d_{23}$ - change 
of the average inter-layer spacing, $b_{1}$,$b_{2}$ - buckling of 
$1^{st}$ and $2^{nd}$ layer, $d_{Ag-O}$ - minimal Ag-O distance, 
$\Delta\Phi$ - oxygen induced work-function change, $\nu$ - 
vibrational frequency perpendicular to the surface. } 
\end{figure}

Starting from the proposed geometry we optimized the structure for 
the missing-row reconstruction (see fig. \ref{FigSketch}). 
However, in the structure proposed on the basis of the XPD 
experiments the oxygen atoms are { 0.15 $\AA$ below the average of 
the surface plane and laterally shifted by 0.36 $\AA$ }towards the 
missing rows  \cite{prb:Rocca:61}, at variance to our results. 
 
{ In the relaxed configuration the oxygen atoms are located at a 
height of about 0.13~\AA~ above the average surface plane in the 
pseudo-hollow sites along the missing-rows; the under-coordinated 
Ag atoms move by $\sim0.2$~\AA~ towards the missing-rows, leading 
to a structure comparable to what 
 is observed for O on Cu(001) \cite{ss:Asensio:236,ss:Kittel:470}. 
Very similar results were obtained in the recent theoretical 
studies by Cipriani et al. \cite{ss:Cipriani:182}and Wang et al. 
\cite{jcp:Wang:106}.} 
 
Besides the differences in structure, the reconstructed phase also 
cannot account for other properties observed for the low 
temperature phase. Experimentally this phase should be 
characterized by an increased vibrational frequency around 
34-37~meV \cite{prb:Rocca:61,el:Benedek:53} and a negative work 
function change \cite{ss:Engelhardt:57}, compared to the 
calculated values of 31~meV for the frequency and a work function 
change of $\Delta\Phi$=+0.5~eV. 
 
From an energetic point of view (surface energy), this 
reconstructed structure is very similar to the simple 
c(2$\times$2) structure. In order to minimize any errors due to 
differences in the setup (k-points, grids for Fourier 
transformations) we recalculated both structures (the 
c$(2\times$2) and the missing row structure) in a 
($2\sqrt2\times\sqrt2$) cell with a denser k-point mesh 
(5$\times$10$\times$1). Using this setup, the surface energy 
(calculated according to equation \ref{surface energy}) for the 
reconstructed structure is even slightly lower in energy 
($\Delta\sigma=$50~mJ/m$^{2}$), so that within the accuracy of our 
calculation both structures are essentially degenerate. 
 
\begin{table} 
\begin{center} 
\begin{tabular}{r|rrr} 
 \hline 
sub-surface oxygens & $\Theta$ [~ML] & $E_{ads}$~[eV] & $\sigma$ [J/$m^{2}$] \\ 
\hline 
   -                                    & 0.50   & -0.94 &  0.09 \\ 
 oct (1)                               & 0.75  & -0.69 &  0.10 \\ 
 tet (2-4)                            & 0.75  & -0.32 &  0.61\\ 
 oct (1),(2)                           & 1.00   & -0.50 &  0.13\\ 
 tet (1-5),(3-1)                      & 1.00   & -0.37 &  0.36\\ 
 oct (1), (2), (3)                     & 1.25  & -0.34 &  0.28\\ 
 tet (6-1), (1-2), (5-4)              & 1.25  &  0.17 &  1.28\\ 
 oct (1), (2), (3), (4)                & 1.50   & -0.14 &  0.59\\ 
 tet  (3-4), (4-6), (1-3), (1-6)      & 1.50   &  0.16 &  1.41\\ 
 \hline 
\end{tabular} 
\end{center} 
\vspace{0.4cm} \caption{\label{TabMR}Several combinations of 
sub-surface oxygen(s) in the missing-row 
($2\sqrt{2}\times\sqrt{2}$) cell in addition to two oxygens in the 
pseudo-hollows (comp. fig. \ref{FigSketch}). The first column 
describes the position of the sub-surface atoms by the atom number 
(according to fig. \ref{FigSketch}) of the nearest on-surface 
atoms. Tetrahedral sites (tet) are between first and second layer, 
octahedral sites (oct) in the second layer. Compiled is the 
resulting total coverage $\Theta$,  the adsorption energy with 
respect to a clean reconstructed surface $E_{ads}$ and the 
generalized surface energy $\sigma$.} 
\end{table} 
 
 Hence, although oxygen adsorption leads indeed to a strong 
weakening of the Ag-Ag bonds, so that structures such as the 
missing-row reconstruction become energetically possible, this 
particular structure is not consistent with the experimental 
findings. A way out of this dilemma would be a more complicated 
structure with additional sub-surface oxygen, as it was already 
argued in Refs. \cite{prb:Rocca:61,el:Benedek:53}. Therefore we 
tested several combinations of occupied octahedral and tetrahedral 
sub-surface sites together with the experimentally proposed 
pseudo-hollow site along the missing-rows (see table \ref{TabMR}). 
However, neither of these structures can either account 
satisfactorily for the observed experimental structure nor is 
energetically more favored than the simple missing row 
reconstruction. 
 
\subsection{Molecular Adsorption} 
Adsorption below 150 K leads to molecular adsorption of oxygen 
\cite{ss:Vattuone:377,bua97b}. { At low coverage two distinct 
molecular species could be identified via vibrational 
spectroscopy\cite{ss:Vattuone:377}. However, since already for 
coverages as low as 0.15 ML one species starts to dominate 
significantly, we will concentrate in the following only on this 
moiety, which saturates finally } in a $c(2\times4)$ structure, as 
observed by STM \cite{mes00}. By comparison with tight binding 
molecular dynamics simulations with these STM results the 
four-fold hollow site was proposed as adsorption site with the 
molecular axis oriented towards the neighboring bridge sites. The 
analysis of thermal desorption spectra \cite{bua97a} lead to an 
estimate for the adsorption energy of 0.4 eV.  This molecular 
species is further characterized  via two dipole active modes at 
84 meV and 30 meV \cite{bua97b} and an O 1s core binding energy of 
around 532.2 eV \cite{prb:Rocca:61} (comp. table \ref{TabExp}). We 
tested several molecular adsorption configurations. The most 
stable position of the molecule is over a hollow site, with the 
molecular axes oriented towards the bridge positions, in agreement 
with the proposed structure from Ref. \cite{mes00}, fig. 4. In the 
$c(2\times4)$ structure (i.e. 0.25 molecules per substrate atom) 
every silver atom binds to only one molecule, so that a quite 
strong binding is possible. The results of our calculation are 
compiled in table \ref{TabO2results}. The molecule is stretched to 
a bond-length of 1.43\AA, situated 1.47\AA~ above the surface 
plane. Similar to atomic adsorption, the first interlayer spacing 
is reduced by $-0.7\%$, whereas the second is slightly expanded 
($+1.1\%$). The magnetic moment of the molecule vanishes 
completely and the work-function of the surface is increased by 
1.83 eV. The calculated adsorption energy (-0.68 eV per molecule) 
is significantly higher than the experimental estimate of -0.4 
eV\cite{bua97a}, reflecting the well known overestimation of 
adsorption energies even with the used gradient corrected exchange 
correlation functionals \cite{ham99}. Also the frequencies are 
overestimated by about 15\%. A possible reason for this could be 
the neglect of substrate movement when calculating the 
frequencies, which could be for a ``soft'' substrate with a low 
melting temperature like silver quite crucial. 
 
\begin{table} 
\begin{center} 
\begin{tabular}{llr} 
\hline 
        \multicolumn{3}{c}{O$_2$ :     $c(2\times4)$ }  \\ 
\hline 
$\sigma$        &$[J/m^2]$      &   0.50    \\ 
$E_{\rm ads}$   &[eV]           &   --0.68    \\ 
$d$             &       [\AA]   &       1.43    \\ 
$h$             &       [\AA]   &       1.47    \\ 
$\Delta d_{12}$ &       [\%]    &       $-0.7$  \\ 
$b_1$           &       [\AA]   &       0       \\ 
$\Delta d_{23}$ &       [\%]    &       $+1.1$  \\ 
$b_2$           &       [\AA]   &       0.06    \\ 
$\Delta \Phi$   &       [eV]    &       1.83    \\ 
$\nu_{\rm O-O}$ &      [meV]    &       97 \\ 
$\nu_{{\rm M-O}_2}$&  [meV]     &       29 \\ 
$\nu_{\rm frust.rot.}$  & [meV] &       39, 38 \\ 
$\nu_{\rm frust.trans.}$ &[meV]&       28, 19 \\ 
\hline 
\end{tabular} 
\end{center} 
\caption{\label{TabO2results} 
Results for molecular oxygen adsorbed in the 
hollow sites of Ag(001) (with the molecular axis pointing towards 
the bridge sites) in a c$(2\times4$) structure. 
The reported quantities are: $\sigma$ - 
surface energy, $E_{\rm ads}$ - adsorption energy per molecule, 
molecular bond length ($d$) and height above the surface plane ($h$), 
 $\Delta d_{12}$, $\Delta d_{23}$ - change of the average 
inter-layer spacing, $b_{1}$,$b_{2}$ - buckling of $1^{st}$ and 
$2^{nd}$ layer and the change of the work-function $\Delta \Phi$. 
 The frequencies ($\nu$) correspond to intramolecular stretch, 
molecule-surface vibration, 
frustrated rotation perpendicular and parallel to the surface, 
and the two frustrated translations parallel to the surface.  } 
\end{table} 
 
\subsection{Silver-oxide - Ag$_2$O} 
 
\begin{figure}[htb] 
 \centerline{ \psfig{figure=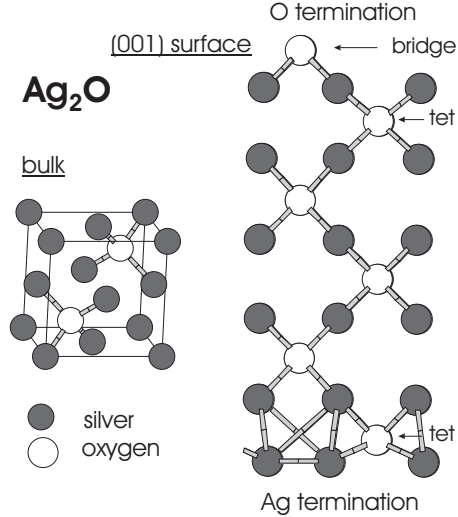,width=6cm,clip=true} } 
\nopagebreak \caption{ \label{Ag2O} Structure of silver oxide 
(Ag$_2$O) in the bulk and of the (asymmetrically terminated) (001) 
oriented surface slab, as it was used in the calculations.} 
\end{figure}

Since it is nowadays well established that on some metals the 
reactive 
 surface is the oxidized one \cite{ove00,hen02}, we performed also a limited set of 
calculations for the (001) surface of Ag$_2$O, which is the stable 
oxide at room-temperature (see Fig. \ref{Ag2O}). The lattice 
constant as determined by the { calculation is $a_{\rm GGA}$=4.83 
\AA ($a_{\rm exp}$= 4.72\AA \cite{cs:Wyckoff:331}). In the DFT 
silveroxide exhibits a vanishing energy gap and becomes hence 
metal-like \cite{deb98} in contradiction to experimental findings, 
according to which Ag$_2$O should be a semiconductor. However, we 
hope that these shortcomings in the electronic description do not 
have a major influence on the results presented in the following. 
 
For the modelling of the surface we} chose a rather small 
super-cell with two silver atoms per layer and a total of six 
Ag-layers as sketched (after relaxation) in Fig. \ref{Ag2O}. With 
this setup we could model at the same time a silver terminated 
surface on one end of the slab and an oxygen terminated one on the 
other side. We fixed the innermost two layers at their ideal bulk 
position and optimized all other layers. The results are compiled in table \ref{TabOxide}. Further results for this 
oxidic surface are given in the subsequent section. 
\begin{table} 
\begin{center} 
\begin{tabular}{llrr} 
\hline 
            &       &   O-terminated &  Ag-terminated   \\ 
\hline 
$\Delta d_{12}$ &       [\%]    &      $-$5.2   &   $-$44.1 \\ 
$\Delta d_{23}$ &       [\%]    &      $-$14.4  &   +3.4    \\ 
$\Delta d_{34}$ &       [\%]    &      $-$5.0   &   +5.8    \\ 
$\Phi$      &       [eV]    &      6.98     &   4.38    \\ 
\hline 
\end{tabular} 
\end{center} 
\caption{\label{TabOxide} 
Results for the (001) surface of silver oxide (Ag$_2$O): 
 Change of the 
inter-layer spacing ($\Delta d$) with respect to the ideal (bulk) Ag-O layer distance and 
the work-function $\Phi$. 
 } 
\end{table} 
 
\subsection{Core Level Binding Energies} 
The phase transition between the low-temperature and high 
temperature phase has been characterized mainly by core-level 
spectroscopy (XPS) \cite{prb:Rocca:61} (comp. table \ref{TabExp}). 
In addition to the O 1s binding energies also the Ag 3$d_{5/2}$ 
core levels have been measured as function of coverage and 
temperature. Hence, we calculated also for some structures the 
binding energies of the core-levels. In these calculations 
vanishing screening was assumed, so that all electrons can 
simultaneously relax into the ground state, after the core 
electron was removed (final state effects) \cite{koe02}. In order 
to minimize errors due to different unit-cells, all calculations 
were performed within a the $p(\sqrt2\times2\sqrt2)$ cell sampled 
by a grid of $6\times3$ k-points. Since the absolute values from 
the calculations are not well defined, only energy differences can 
be compared. For silver the bulk core level (calculated with in 
the same super cell) is an appropriate reference, for oxygen we 
chose hollow adsorption at low coverage ($\Theta=0.25$ ML). 
 
\subsubsection{Ag $3d$} 
\label{SecCoreAg} 
\begin{figure}[htb] 
 \centerline{ \psfig{figure=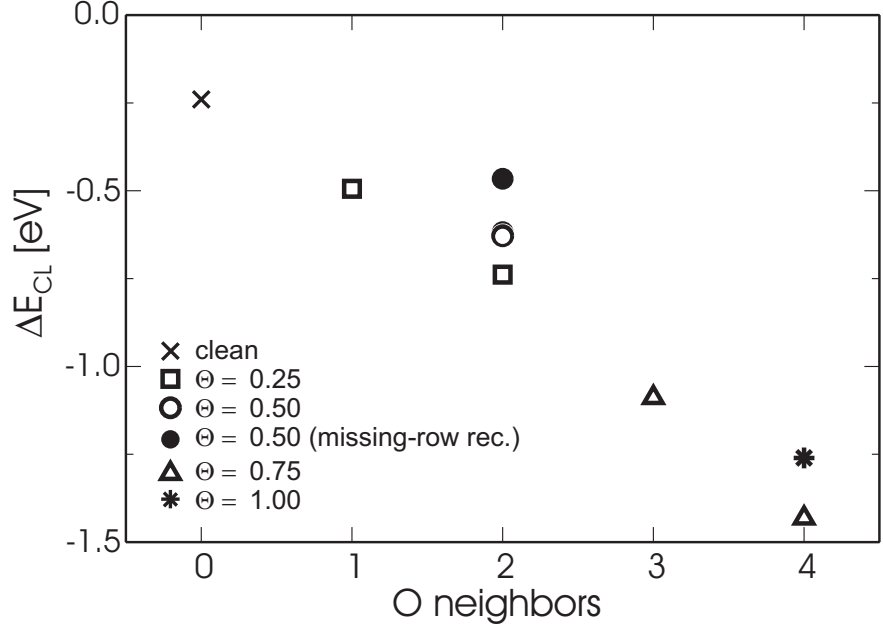,width=12cm,clip=true} } 
\nopagebreak \caption{ \label{FigCore} Ag 3d surface core-level shifts 
 as a function of the number of 
oxygen neighbors for different coverage. 
All values have been calculated in a p$(2\sqrt2\times\sqrt2)$ cell. 
The binding energy of bulk silver calculated in a comparable cell was taken as reference.} 
\end{figure}

Experimentally, for the Ag 3$d_{5/2}$ core level 
only two peaks could be resolved independent of temperature and 
oxygen coverage: One peak at 368.0 eV for the clean surface 
component and a second at 367.6 eV for silver atoms in contact 
with oxygen.{ \cite{prb:Rocca:61} } While for the low temperature 
phase both peaks are almost of equal intensity, the surface 
component clearly dominates at high temperatures. Fig. 
\ref{FigCore} summarizes the results of our calculations. On the 
clean surface the Ag 3$d_{5/2}$ energy is shifted by approximately 
240 meV with respect to the bulk Ag core-level. With increasing 
oxygen coverage the binding energy shifts down, almost linearly 
with the number of oxygen neighbors, as it was observed already 
for oxygen on Ru(0001) and Rh(111) \cite{liz01}. Since already a 
single oxygen neighbor induces a shift (with respect to the clean 
surface) of about 0.25 eV, the experimentally observed shift of 
$-0.4$ eV implies that there are no silver atoms with more than 
two oxygen neighbors. This is also in line with the experimental 
interpretation 
 of a maximum coverage of 0.4 ML \cite{prb:Rocca:61}. Slightly different is the situation 
for  the missing row reconstruction. Here the reduced number of 
silver nearest neighbors shifts the core-level up by approximately 
150 meV, thus compensating partially for the oxygen induced shift 
(comp. fig. \ref{FigCore}, full symbols). 
 
\subsubsection{O $1s$} 
\begin{figure}[htb] 
\centerline{\psfig{figure=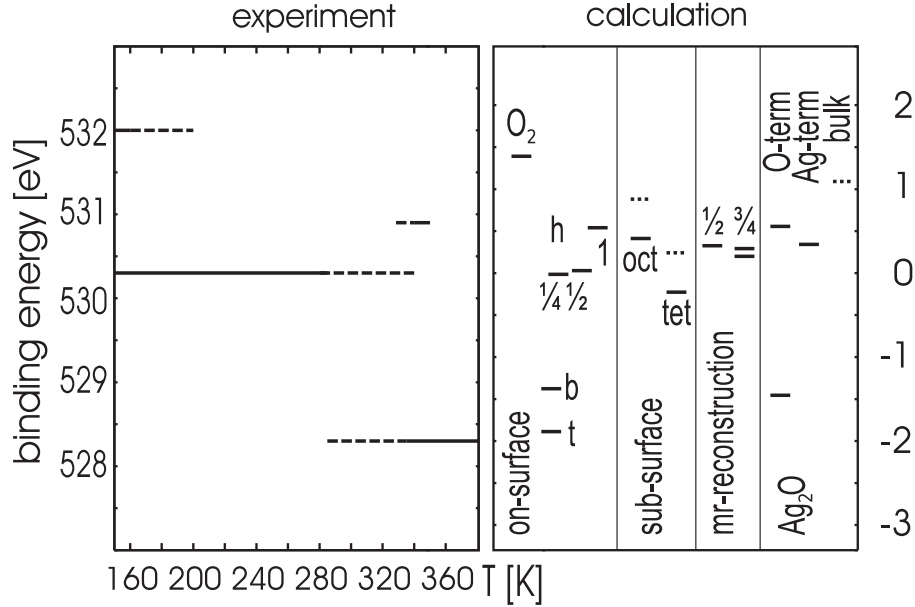,width=12cm,clip=true}} 
\nopagebreak \caption{\label{FigExpCore} Experimentally determined 
O 1s binding energies for varying temperature, taken from 
Ref.\cite{prb:Rocca:61} (left panel). Calculated core-level 
binding energies for different coverages and adsorption geometries 
(right panel); numbers indicate the coverage in mono-layer, dashed 
lines refer to oxygen in bulk positions.} 
\end{figure}

The experimental results for the oxygen 1s core-levels { from 
Rocca et al. \cite{prb:Rocca:61} } are more complicated (comp. 
table \ref{TabExp}, fig.\ref{FigExpCore}): At lowest temperature 
the molecular species can be identified via its binding energy 
around 532.0 eV. Dissociation of the molecules leads to a 
prominent peak at 530.3 eV up to a temperature of around 300K, 
where this peak starts to vanish. Prior to disappearing another 
small peak appears at approximately 530.9 eV, which could have 
been there already before, but buried by the main peak at 530.3 
eV. This peak vanishes also at 350 K. Above 300K the 
high-temperature species gives rise to a peak at around 528.3 eV. 
 
Table \ref{TabOCore} summarizes the results for various structures 
and adsorption sites. All energies are given with respect to the O 
1s binding energy of hollow-adsorbed oxygen at low coverage 
($\Theta=0.25$ ML). The core-level is mainly determined by the 
coordination of the adsorption site and changes hardly with 
coverage, as can be seen from the comparison of the core-level for 
hollow adsorbed oxygen at $\Theta=0.25$ and 0.50 ML. Only for 
$\Theta=1.00$ ML the binding energy is larger, since the special 
adsorption geometry slightly below the surface layer for this 
coverage leads to a higher ``effective coordination number''. Also 
the values for the missing row reconstructed surface have slightly 
higher binding energies compared to hollow adsorption. Again the 
low adsorption height increases the effective coordination and 
hence also the O 1s binding energy. The subsurface sites continue 
this trend: the oxygen binding energy of the lower coordinated 
tetrahedral site is 0.2 eV below that of the hollow adsorbed 
species, for the higher coordinated octahedral site the binding 
energy increases even more. The highest binding energy is 
calculated for the molecular state: in addition to the 
high-coordinated adsorption site there is a substantial 
oxygen-oxygen bond. The lower coordinated bridge and on-top 
adsorbed atoms continue the trend at the other end of the 
spectrum.

\begin{table} 
\begin{tabular}{lllr} 
\hline 
$\Theta_{\rm O}$        &       cell    &       site(s)&        $\Delta E_{\rm CL}$(O 1s) [eV]  \\ 
\hline 
0.5                     &       $c(2\times4)$O$_2$              &       h       &               $+1.40$\\ 
0.25                    &       $p(\sqrt2\times2\sqrt2)$        &       h       &           $\equiv$0.00 \\ 
0.25                    &       $p(\sqrt2\times2\sqrt2)$        &       b       &           $-1.36$\\ 
0.25                    &       $p(\sqrt2\times2\sqrt2)$        &       t       &           $-1.88$\\ 
0.25                    &       $p(\sqrt2\times2\sqrt2)$        &       tet     &           $-0.20$\\ 
0.25                    &       $p(\sqrt2\times2\sqrt2)$        &       oct     &           $+0.45$\\ 
0.50                    &       $p(\sqrt2\times2\sqrt2)$        &       $2\times$h      &       $+0.01$\\ 
1.00                    &       $p(\sqrt2\times2\sqrt2)$        &       $4\times$h      &       $+0.61$\\ 
\hline 
0.50                    &mr-$p(\sqrt2\times2\sqrt2)$    &       $2\times$h      &               $+0.39$\\ 
0.75                    &mr-$p(\sqrt2\times2\sqrt2)$    &       $2\times$h+oct  &               $+0.35/+0.21$\\ 
\hline 
$-$                    &bulk-$p(\sqrt2\times2\sqrt2)$  &       tet          &           $+0.22$\\ 
$-$                    &bulk-$p(\sqrt2\times2\sqrt2)$  &       oct          &           $+0.91$\\ 
$-$                    &bulk Ag$_2$O-$p(\sqrt2\times2\sqrt2)$   & $-$       &           $+1.07$\\ 
\hline 
$-$                    &Ag$_2$O-$p(\sqrt2\times\sqrt2)$-O-term  &       b/tet   &           $-1.45/+0.57$\\ 
$-$                    &Ag$_2$O-$p(\sqrt2\times\sqrt2)$-Ag-term  &      tet     &       $+0.35$\\ 
\end{tabular} 
\caption{\label{TabOCore} Oxygen 1s core level energies for oxygen adsorbed for various coverage, sites (h- 
hollow, b-bridge,t-top,tet - tetrahedral subsurface site, oct - octahedral subsurface site)  and structures. All 
energies are given with respect to the binding energy of the hollow adsorbed species at low coverage.} 
\end{table}

\subsection{Optimal adsorption sites with increasing coverage: mixed surface and subsurface adsorption} 
\begin{figure}[htb] 
\centerline{\psfig{figure=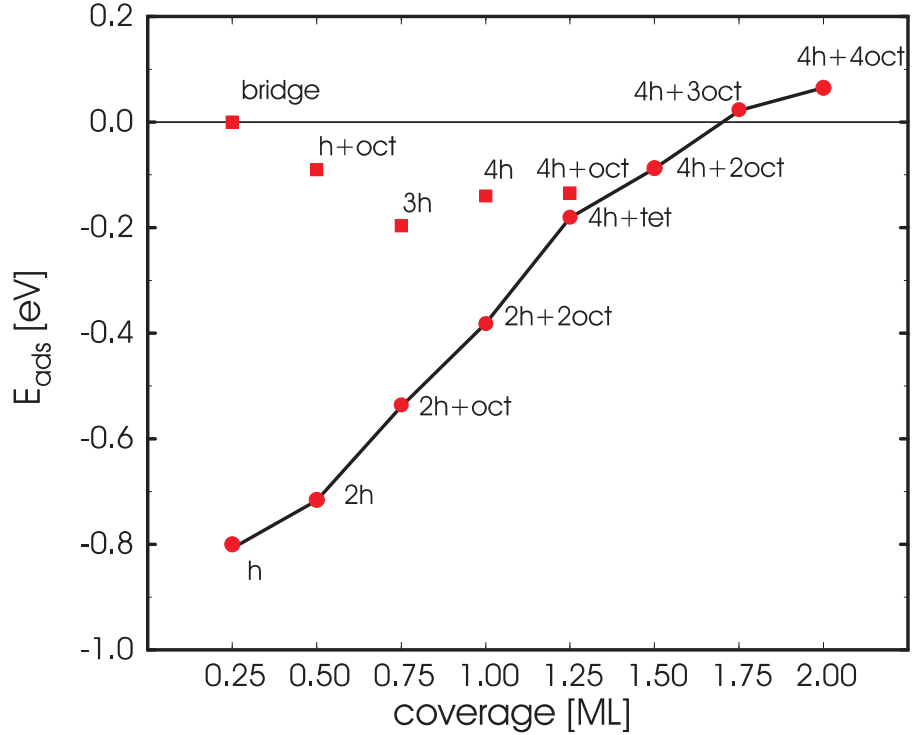,width=12cm,clip=true}} 
\nopagebreak \caption{\label{FigCov} Adsorption energy $E_{ads}$ 
for the structures at the coverage from 0.25~ML to 2.0~ML 
calculated in a 
 p$(2\times2$) cell. The second favorite structure is 
indicated by squares. } 
\end{figure}

 To complete our study we investigated several more 
adsorption structures (in total more than 50 different structures) 
with mixed on-surface and sub-surface oxygen for total oxygen 
coverage between $\Theta=0.11$~ML and $\Theta=2$~ML. The 
adsorption energy for a selection of adsorption structures as a 
function of oxygen coverage is plotted in fig. \ref{FigCov}. Up to 
half monolayer coverage the hollow site is the most favorable 
adsorption position; at $\Theta=0.5$~ML only the missing-row 
reconstruction is degenerate with the c($2\times2$) structure 
(comp. Sec. \ref{Missing-row reconstruction}). At higher coverage 
combinations of sub-surface and on-surface sites become more 
stable: at $\Theta=0.75$~ML a combination of two hollow and one 
octahedral sites in a p($2\times2$) cell, at full coverage half of 
the oxygen is located below the surface in octahedral sites and 
half of it on the surface in the hollow sites. Interestingly for 
adsorption in isolated sub-surface sites, the tetrahedral site is 
more stable (see fig. \ref{FigPath}). This is 
also true in combination with 1~ML of on-surface oxygen (see fig. 
\ref{FigCov}), but the octahedral site becomes clearly more 
favorable for higher sub-surface coverage. 
 
\section{Summary and Discussion} 
 
\subsection{Surface energy} 
\begin{figure}[htb] 
\centerline{\psfig{figure=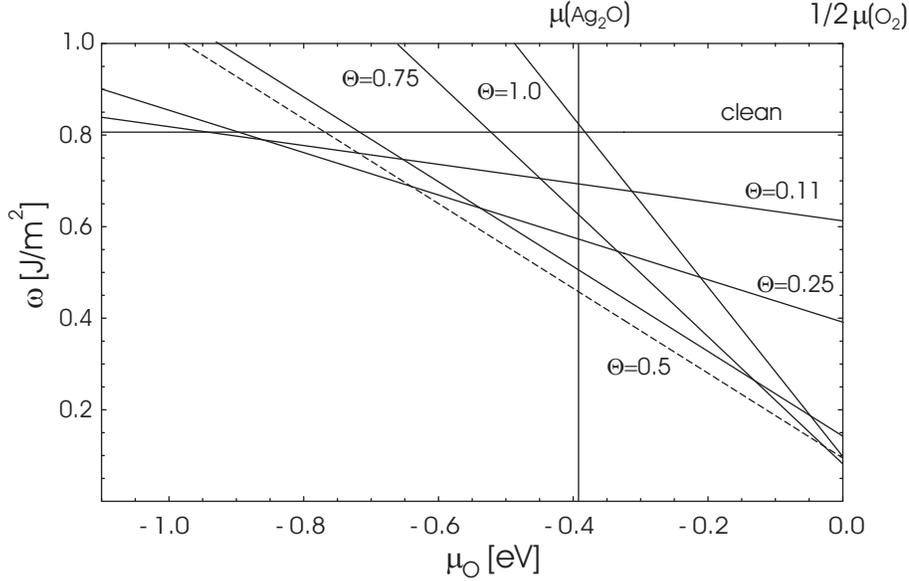,width=12cm,clip=true}} 
\nopagebreak \caption{\label{FigFavourable} Adsorption energy per 
area with respect to the chemical potential. Only the most 
favorite 
 structures for each coverage (comp. fig. \ref{FigCov}) are shown). 
Additionally included is the half coverage missing row structure 
with a dashed line. 
$\frac{1}{2}\mu(O_{2})$~=~0~eV corresponds to free molecular 
oxygen and $\mu(Ag_{2}O)$~=~-0.39~eV to silver-oxide.} 
\end{figure} 
 
In order to facilitate a direct comparison of all structures with a 
different number of Ag atoms in the surface (missing row 
reconstruction, silver oxide Ag$_{2}$O, $\dots$), we use the (surface) 
 grand potential, i.e. the Legendre 
transform of the generalized surface energy (Eqn. \ref{surface 
energy})  with respect to the number of oxygen atoms 
\begin{equation} 
\omega(\mu_O)=\sigma-\frac{N_O}{A} \mu_O \end{equation} where 
$\mu_O$ stands for the chemical potential of oxygen (with 
$\mu_O(O_2)\equiv0$ eV).{  Via this transformation the dependence 
on the number of (oxygen) particles is transformed into a 
dependence on the chemical potential. So every structure is 
represented by a line, where the intercept at the ordinate equals 
the surface energy and the slope increases with coverage. In Fig. 
\ref{FigFavourable} this quantity is plotted. 
 It describes the stability of different Ag--O 
 structures as a function of varying oxygen chemical potential. For fixed temperature (partial pressure) 
 the chemical potential scale is equivalent to partial pressure (temperature). 
  A more detail description can be found in the paper by Reuter and Scheffler 
  \cite{prb:Reuter:65}.} 
  In the limit of low oxygen exposure (partial 
pressure, chemical potential) on the left hand side of the plot 
 the clean silver surface exhibits the lowest grand potential and is hence the thermodynamically 
 most stable structure. 
When the chemical potential increases to around { $-0.95$ eV the 
thermodynamical equilibrium moves} to the lowest investigated 
oxygen coverage ($\Theta=\frac19$ ML). With increasing chemical 
potential 
 the $p(2\times2)$ structure at a quarter coverage becomes stable, 
followed by the two very similar half coverage structures 
($c(2\times2)$ and missing row reconstruction). The calculated 
chemical potential of oxygen in the Ag$_2$O silver oxide structure 
(i.e. the heat of formation of the oxide) of $-0.39$ eV 
(experimental value: $-0.32$ eV \cite{weast}) represents the 
oxygen-rich limit for the investigated structures. In other words: 
for oxygen chemical potentials above $-0.39$ eV it is (from a 
thermodynamical point of view) more favorable to convert the 
silver surface into Ag$_2$O, than increasing the coverage at the 
surface beyond 0.5 ML.

The result of a maximum oxygen coverage of around 0.5 ML is 
consistent with the experimental estimate of a maximal coverage of 
around 0.4 ML\cite{prb:Rocca:61}. Higher oxygen exposure leads 
already to an oxidation of the silver surface. This is also in 
line with the calculated core level binding energies. 
 
\subsection{Core-level energies} 
In section \ref{SecCoreAg} we demonstrated that there is a linear 
dependence between the Ag core level binding energy and the number 
of oxygen neighbors. The comparison with the experimentally 
determined core-level shift of $\sim$0.4 eV \cite{prb:Rocca:61} 
implies that most silver surface atoms have only one oxygen 
neighbor. Only if also reconstructions, which reduce the 
coordination of the surface silver atoms are taken into account 
also two oxygen neighbors become possible (comp. fig. 
\ref{FigCore}). The results for the oxygen 1s binding energies are 
less simple to interpret. 
 
Figure \ref{FigExpCore} summarizes once more the experimental 
results. The main (low temperature) species is the hollow adsorbed 
oxygen species with a coverage less than 0.5 ML, which marks also 
the reference energy of the results compiled in table 
\ref{TabOCore}. The molecular species is characterized by a higher 
binding energy; $\Delta E_{\rm CL}\cong$1.4 eV in our calculation 
compared to 1.7 eV in Ref. \cite{prb:Rocca:61}. This discrepancy 
might be attributed to coverage effects for both the molecular and 
the atomic adsorbates. With increasing temperature Ag vacancies 
form and parts on the surface will undergo an oxygen-induced 
missing row reconstruction; perhaps with additional oxygen in 
residing in subsurface sites. Increasing the temperature even more 
the mobility within the surface layer also increases and the 
(experimentally observed) phase transition sets in, leading to two 
new peaks in the XPS signals at around 528.3 eV and 530.9 eV, i.e. 
at $-2.0$ eV and +0.6 eV with respect to the low temperature peak. 
The peak at $-2.0$ eV could indicate low-coordinated oxygen 
species (bridge, on-top sites). However, the energy difference 
between hollow and top sites is so large, that an occupation of 
on-top sites is highly unlikely at ambient temperatures. Bridge 
sites represent the saddle points for a diffusion between hollows 
- the calculated core-level shifts, however, are too small to 
account for the observed low-energy peak around 528.3 eV. It is 
more realistic to assume that the phase transition finally leads 
to the formation of an (oxygen terminated) oxide layer on top of 
the silver surface, characterized by oxygen atoms in bridge and 
tetrahedral interstitial sites of the silver substrate. The 
calculated values of $-1.45$ eV and $+0.57$ eV for the perfect 
silver oxide layer agree reasonably well with experimental values 
($-2.0$ and $+0.6$ eV), considering that the newly formed oxide is 
never perfectly flat and will not only consist of (100) facets, 
but also other surface orientations. Furthermore, the lattice 
constant will be determined by a compromise between that of the 
underlying silver substrate { $a_{\rm exp}$=4.08\AA\ ($a_{\rm 
GGA}$=4.16\AA) and that of Ag$_2$O $a_{\rm exp}$=4.72\AA\ ($a_{\rm 
GGA}$=4.83\AA).}

\subsection{Work-function and frequencies} 
The results for work function and stretch frequencies provide 
further information. However, we are unable to explain the 
negative work-function change observed at low coverage and 
temperatures in Ref. \cite{ss:Engelhardt:57}. The pronounced 
increase of the work-function with temperature and coverage is 
consistent with the higher values for $\Delta \Phi$ for the lower 
coordinated sites. 
 
The adsorbate-substrate stretch frequencies for this particular 
system also do not permit a unique identification of the 
adsorption geometry. Both, the experimental and the calculated 
values, are biased by high uncertainties. On the experimental side 
the unknown coverage and different preparation methods lead to 
values differing by about 10 \%. For the calculation the strong 
coupling between soft surface modes of the substrate with the low 
vibrational frequencies of a relatively heavy adsorbate complicate 
the analysis. This mode coupling could only be taken into account 
by performing full ab-initio molecular dynamics simulations over 
long time intervals. 
 
\subsection{Surface mobility} 
The high mobility of the silver surface is an important ingredient 
for its complicated behavior. Hence we want to spend a few words 
on this topic. During our study we performed additionally several 
ab-initio molecular dynamics simulations, mainly for simulated 
annealing optimizations for various structures and surface 
stoichiometries. E.g. by a simulation within a $p(2\sqrt2\times2)$ 
cell with a surface stoichiometry of Ag:O=6:3, we were able to 
reproduce the missing row reconstruction. An interesting byproduct 
of these simulations is the very high mobility of the surface 
silver atoms, induced by the strong weakening of Ag-Ag bonds due 
to the formation of strong Ag-O bonds. Although we cannot give a 
quantitative description of this behavior, in all our simulations 
the mobility of the silver atoms was much higher compared to that 
of the oxygen atoms, which were hardly leaving their hollow sites. 
This observation is further supported by STM observations of a 
high mobility of Ag(100) surfaces under atmospheric conditions at 
room temperature\cite{wic92}. This mobility favors the formation 
of vacancies and other defects well below the melting temperature{ 
, so that the effective silver coordination of the oxygen atoms is 
reduced}. The consequence is an early onset of the metal-oxide 
transition accompanied by a variety of defects and small facettes 
exhibiting different orientations of the metal as well as the 
oxide surface. All this complicates the comparison with well 
defined surface structures as investigated in a theoretical study. 
 
\section{Conclusion} 
\label{Conclusion} We have presented detailed ab-initio 
density-functional studies of the atomic structure, energetics, 
and other properties of the clean and oxygen covered Ag(001) 
surface. Besides adsorption in high symmetry positions also 
molecular adsorption, an oxygen-induced missing-row reconstruction 
and the Ag$_2$O silver oxide surface was investigated. In 
comparison with data taken from the experimental literature we 
find the following scenario: At low temperature oxygen adsorbs 
molecularly, forming a $c(2\times4)$ over layer of molecules 
residing in hollow sites. At about 200K these molecules 
dissociate. The atoms adsorb at hollow sites with coverages never 
exceeding 0.5 ML. Due to the strong Ag-O bond, the intermetallic 
bonds are weakened and partially broken, so that vacancy 
structures such as the missing row structure become possible. At 
even higher temperatures (around 320 K), all atoms at the surface 
become mobile, so that during silver diffusion the coordination 
number of the oxygen adsorption sites is reduced to 3 or even 2. 
With increasing oxygen exposure and temperature the silver surface 
turns into an Ag$_2$O surface covered with bridging oxygen atoms. 
By comparison with experimental findings\cite{prb:Rocca:61}, we 
propose that this species is finally the reactive one for e.g. CO 
or C$_2$H$_4$ oxidation. 
 
\section*{Acknowledgement} 
We want to thank L.K\"ohler and G. Kresse for the implementation of core-levels in VASP and for providing a 
preliminary version.

\end{document}